\documentstyle[11pt,aaspp4]{article}

\begin{document}

\title{The X-ray Luminosity Function and Gas Mass Function for Optically-Selected
Poor + Rich Clusters of Galaxies} 
\author{Jack O. Burns, Michael J. Ledlow,  Chris Loken, \& Anatoly Klypin} 
\affil{Department of Astronomy, New Mexico State University,  \\ Las Cruces, NM
88003-0001 \\ jburns, mledlow, cloken, aklypin@nmsu.edu}
   
\vskip 0.3in
\author{Wolfgang Voges}
\affil{Max-Planck-Institut f\"ur Extraterrestrische Physik \\ Postfach 1603, D-85740, Garching
bei M\"unchen, Germany \\ whv@mpe.dnet.nasa.gov}

\vskip 0.3in
\author{Greg L. Bryan \& Michael L. Norman}
\affil{Department of Astronomy \& National Center for Supercomputing Applications \\ 5600
Beckman Institute, Drawer 25 \\ University of Illinois at Urbana-Champaign, Urbana IL
61801 \\ gbryan, norman@ncsa.uiuc.edu}

\vskip 0.3in 
\author{Richard A. White}
\affil{Space Data and Computing Division,  Code 932 \\ NASA/Goddard Space Flight Ctr.,
Greenbelt, MD 20771 \\ rwhite@jansky.gsfc.nasa.gov}

\vskip 0.5in
\begin{center}
To appear in the {\bf Astrophysical Journal Letters}.
\end{center}

\begin{abstract}
We present the first X-ray Luminosity Function (XLF) for an optically-selected
sample of 49 nearby poor clusters of galaxies and a sample of 67 Abell clusters
with $z \le 0.15$. We have extended the measured cluster XLF by more than a factor of 10
in X-ray luminosity. Our
poor cluster sample was drawn from an optical catalog of groups with
$0.01 \le z \le 0.03$ composed of Zwicky galaxies.  The X-ray emission was measured
from the ROSAT all-sky survey. About 45\% of the poor clusters were detected with
0.5-2.0 keV luminosities from $(1.7-65) \times 10^{41} h^{-2}$ ergs/sec.  These are 
among the X-ray brightest, optically-selected poor clusters in the northern hemisphere.
For this sample, the poor cluster XLF was found to be a
smooth extrapolation of the rich cluster XLF.  A new Hydro/N-body simulation of a Hot +
Cold dark matter model with $\Omega_{total}$=1, $\Omega_{\nu}$=0.2, and a baryon
fraction of 7.5\% was used to model and understand our observational
selection effects.  We found that the observed cluster Gas Mass Function 
was consistent with our model.
\end{abstract}


\section{Introduction}

The presence of head-tail radio galaxies (e.g., Venkatesan et al. 1994) and extended X-ray
emission (e.g., Ebeling et al. 1994; Doe et al. 1995; Pildis et al. 1995) demonstrate the
existence of a relatively dense ($\approx$5$\times 10^{-4}$ cm$^{-3}$) and hot ($\approx$1 keV)
intracluster medium within poor clusters. The X-ray emission, in particular, is a good tracer of the
gravitational potential well and the dynamical state of the clusters.  {\it Einstein}
observations (Price et al. 1991) showed that the bulk X-ray and
optical properties of these poor clusters appear to be scaled-down versions of the rich clusters,
a result consistent with {\it ROSAT} observations (Doe et al. 1995). 
However, recent work (e.g., Mulchaey et al. 1996), including new ASCA results,
suggests that poor clusters may have a wide range of baryonic fractions (5-30\%) and low
metallicity ($<0.15$ solar, Davis et al. 1996; Tawara et al. 1995), which might indicate that
they are distinctly different from rich clusters.  We are only beginning to understand the
formation and evolution of poor clusters; models of their genesis remain controversial (e.g.,
Diaferio et al. 1993).
  
In an effort to investigate the properties of poor galaxy clusters and to provide a
reference point for studies of cluster evolution, we have constructed an X-ray Luminosity
Function (XLF) for an optically-selected, nearby, volume-limited sample using images from the {\it
ROSAT} all-sky survey (RASS).  Previous XLFs have concentrated mainly on rich
clusters (e.g., Edge et al. 1990; Henry et al. 1992; Castander et al. 1994), although Henry et al. (1995) have
recently added one low $L_X$ point to the XLF from a small sample of X-ray selected groups.
In this paper, we also calculate the Gas Mass Function for
poor and rich clusters, and compare this function with that computed from large-scale structure
models.  We use $H_{\circ}$ = 100 $h$ km/sec/Mpc and $q_{\circ}=0.5$ throughout.
   
\section{Poor Cluster Statistical Sample \& RASS Images} 

Using an algorithm similar to that of Turner \& Gott (1976), White et al.
(1996) compiled a catalog of $\approx$600 optically selected poor clusters composed of
Zwicky et al. (1961-68) galaxies down to $15\fm7$.  These groups were identified by drawing
the largest possible circle, centered on a Zwicky galaxy, within which the surface density
of Zwicky galaxies was $>$24 times the average surface density in the north galactic
cap.  This catalog contains nearly all the Yerkes poor clusters (e.g., White 1978),
a few Hickson groups (4 out of 49 clusters in Table 1), as well as many more loose
condensations.  A partial listing of the densest poor clusters in the catalog along with
VLA observations were published by Burns et al. (1987).  For this X-ray project,
we selected a complete subsample of the 49 densest poor clusters, as defined in Ledlow et al. (1996), 
each with (a) $\ge$4 Zwicky galaxies, (b) $\mid b \mid>30\arcdeg$, (c) a surface overdensity
$\gtrsim$50, and (d) $0.01\leq z \leq 0.03$.  As a result, the effective radii
of the groups in Table 1 range from 60 to 180 $h^{-1}$ kpc for $z=0.01$ to
$z=0.03$, respectively.  Eight Abell clusters also lie within this volume (see Ledlow et al. 1996).

The fact that the volume density of our poor clusters is constant within this restricted
redshift range (Ledlow et al. 1996) indicates that our sample is basically volume-limited.
Our apparent magnitude cutoff reduces the volume density of observable galaxies by a factor of $\sim$5
across our volume, whereas the projected area within which the candidate group galaxies
must lie increases by a factor of 9 due to the surface overdensity condition.  Therefore,
it appears plausible that these two effects cancel, at least to within a factor of 2, in our volume.

We cross-correlated this sample of poor clusters with the RASS (Voges 1992).  {\it ROSAT} was nearly
ideal for imaging groups since the PSPC detector was most sensitive to cooler (1 keV) clusters.  
From the RASS sky scans, we assembled $2\fdg1 \times 2\fdg1$ images from 0.5-2.0 keV with
average exposure times of $\approx$550 sec, corresponding to an X-ray flux limit of
$\approx$4$\times 10^{-13}$ ergs/cm$^2$/sec. After exposure and background
corrections, the images were smoothed with $204\arcsec$ FWHM Gaussians ($\approx$60 $h^{-1}$
kpc at $z=0.02$) to accentuate the extended ICMs.  Optical identifications of the X-ray sources
were made by overlaying the RASS maps onto digital POSS-I images.  Examples of RASS X-ray
emission for 4 poor clusters are shown in Fig. 1.
 
For those clusters with X-ray detections, we summed the X-ray counts within a circle of linear
radius 190 $h^{-1}$ kpc about the centroid.  We fit the surface brightness profiles
of the 7 best cluster detections with a $\beta=2/3$ King model convolved with the smoothed
PSF determined from point sources in the fields.  We measured an average $r_c$=65
$h^{-1}$ kpc which was used in correcting the flux within 190 $h^{-1}$ kpc to a total
flux (following Briel \& Henry 1993); this correction was 67\%.  The resulting RASS 0.5-2.0 keV
luminosities along with $3\sigma$ upper limits for undetected clusters are given in Table 1.
We include a 10\% calibration error in $L_X$ added in quadrature
with the Poisson error in the X-ray counts.  The conversion between count rate and luminosity was
performed using XSPEC assuming a Raymond-Smith thermal spectrum with T=1 keV, 0.3 solar
abundance, and HI column densities from Stark et al. (1992).
  
A total of 22/49 (45\%) of the clusters in the sample were detected by the RASS. This X-ray
detection rate is higher than that recently reported for other galaxy groups (e.g., 
Ebeling et al. 1994; Mulchaey et al. 1996).  These detections represent some of the X-ray brightest, optically-selected poor clusters in the northern sky.  The X-ray
luminosities span the range $(1.7-65)\times 10^{41} h^{-2}$ ergs/sec, which
run the gamut from that expected for bright, individual galaxies to richness class
$\gtrsim$0 Abell clusters.  We do not include unresolved X-ray sources associated with individual galaxies as
cluster detections.  However, it is possible that some fraction of the total X-ray emission for
these extended sources is due to emission from galaxies (e.g., AGNs, ISMs or interacting
galaxies; see e.g., Ebeling et al. 1994).  From the log $N$ - log $S$ relationship for the RASS, we
expect a total of only 6 random source projections within the search radii of these 49 clusters; but,
this is an upper limit since we eliminated compact and noncluster sources using optical IDs.
 
Next, we ask how many of the poor clusters in our sample may be artificial systems produced by
projection effects?  Given the complex observational selection effects inherent in our sample,
we addressed this question using a new N-body + Eulerian Hydro large-scale structure simulation (Bryan
\& Norman 1996; Loken et al. 1996).  The simulation was done for a mixed Cold + Hot (two
neutrino) dark matter (CHDM) model (Primack et al. 1995) in a (50 $h^{-1}$ Mpc)$^3$ box with a
$512^3$ mesh (98$h^{-1}$ kpc/zone), and $3\times256^3$ particles. The model assumed $h$=0.5,
$\Omega_{CDM}$=0.725, $\Omega_{\nu}$=0.20, $\Omega_{b}$=0.075, and $Q_2$=19 $\mu$K, and was
run on the CM-5 at NCSA.  Local maxima in the cold dark matter distribution were identified as
``galaxy halos'' and periodic boundary conditions were used to replicate the galaxies into a
larger volume (see Loken et al. 1996 for details).  A vantage point was chosen for an observer, and a
Schechter selection function (e.g. Klypin et al. 1990) was applied to statistically select a catalog of
Zwicky-like galaxies. We then identified groups in the same manner as discussed at the beginning
of this section.  We found $\approx$58 clusters (there is some variation with vantage point) in a volume with $0.01<z< 0.03$.   Roughly (80-86)\% of the groups had 4 or more galaxies within $2 h^{-1}$ Mpc
and virtually all ($>96$\%) had at least 2 such members. Moreover,
(88-98)\% of the clusters selected from this 2D percolation algorithm were spatially
coincident with real clusters which we had previously identified in the 3D volume.
We conclude that our poor clusters are nearly all likely to be real and not produced by
projection effects.

 \section{The Cluster X-ray Luminosity Function}

The differential XLF for our poor clusters is represented by the open circles in Fig. 2.
Poisson error bars are plotted and these points include the 8 Abell clusters found in the same
volume as our poor clusters.  The X-ray observations are surface-brightness-limited and contain a small
Malmquist-type bias; so, the first point in the XLF had to be computed using the smaller
volume to which the first bin is complete to compensate for the bias.

We also plot the XLF for rich clusters as filled circles in Fig. 2 using a subsample of
Abell clusters observed with the RASS by Briel \& Henry (1993).  We used only those
clusters with $z \le 0.15$, which is a relatively complete sample of richness class $\ge$0
clusters (e.g., Ebeling et al. 1996).  Luminosities were corrected to our cosmology, and
converted to our observed energy band assuming a free-free spectrum with $T=3$ keV and
$r_c=95 h^{-1}$ kpc for richness class 0 clusters, and $T=6$ keV and $r_c=125 h^{-1}$ kpc
for all others.  Our subsample contains 67 Abell clusters of which 33 (49\%) were detected
with luminosities ranging from $(0.3-8.1) \times 10^{43} h^{-2}$ ergs/sec.  The first two
rich cluster bins also suffer from a Malmquist bias and were corrected in the same manner
as the poor clusters.

In computing the XLF, we used both a direct binning approach (treating upper limits as
detections) and a Kaplan-Meier (e.g., Feigelson \& Nelson, 1985) estimator for detections
and upper limits.  The two methods produced identical slopes for the differential and
cumulative XLFs.

As shown in Fig. 2, the XLF is well-fit to a power-law in the form: $-5.98(\pm0.22) - 1.71(\pm0.19)\times
log(L_X/10^{44})$.  Within a 90\% confidence level, our XLF agrees with those of Edge et al.
(1990) and Henry et al. (1995) in regions where they overlap.  It appears that the XLF for
our optically-selected poor clusters is a smooth extrapolation of the XLF for nearby Abell
clusters.  This further suggests that the bulk X-ray properties of poor clusters are not fundamentally
different from rich clusters.

\section{The Intracluster Medium Mass Function}

We estimated the gas masses for the poor clusters in our sample using the measured $L_X$
values and again assuming that X-ray surface brightness follows a $\beta$-model.
The X-ray mass within a radius
$r$ is given by $M_{gas}(\le r) = 4 \epsilon^{-1/2} L_X^{1/2} r_c^{3/2} [r/r_c - {\arctan
(r/r_c)}]$ for $\beta=\twothirds$, where $\epsilon$ is the volume emissivity (calculated for
a Raymond-Smith plasma using XSPEC).  In Table 1, we list the gas masses within $1.5 h^{-1}$ Mpc of the cluster
centers assuming $r_c=65 h^{-1}$ kpc and $T=1$ keV.  

In an effort to assess the reliability of our observational gas mass estimates, we calculated
the gas masses of our 3D clusters in the CHDM volume using the above approach based on their luminosity,
and compared this with their known gas masses.  We find that the observational mass determination
does a reasonable job of estimating the true masses although there is substantial scatter ($\approx$50\%
standard deviation).
  
In Fig. 3, we show the integral Gas Mass Function (GMF) for the
cluster samples.  We also calculated the GMF that would be observed in the CHDM simulation volume. The GMF is
somewhat more reliable in grid-based numerical simulations than the XLF (Anninos \& Norman 1996).
Since we had previously compiled a list of 3D X-ray clusters within the 
volume, we were able to correlate their projected positions on the sky with those of the
optical groups. We found that (77-85)\% of the optical groups had projected X-ray emission within
$20\arcsec$ of the optical group centroid.  The GMF for those 3D clusters which coincided with a complete,
volume-limited subsample of our optical groups using the same selection
function as the observed groups is shown in Fig. 3.  The fairly good agreement between
the model and the observations for the poor clusters again suggests that our
observed clusters in Table 1 form a nearly complete subsample of all groups with the
selection criteria given in $\S$2.

The curves in Fig. 3 correspond to Press-Schechter (PS) predictions for the
GMF in 3 different cosmologies. The PS GMF for the CHDM model agrees well with
the GMF for the Abell clusters but lies above that of the poor clusters.  This
is not unexpected because of the particular selection criteria used in constructing
the sample in Table 1; we have selected only a subset (albeit complete) of all possible groups within our
observed volume. The remaining two curves correspond to flat, COBE-normalized, $\Lambda$+CDM
models ($Q_2=21.8 \mu$K): one with $\Omega$=0.5, $\Omega_b$=0.035, $\sigma_8$=1.22, $h=0.6$ and the
other with $\Omega$=0.3, $\Omega_b$=0.03, $\sigma_8$=1.09, and $h=0.7$. We see that the GMF
amplitude drops as the total mass density of the universe decreases and that the
GMF can potentially be used to constrain cosmological models. The $\Lambda$+CDM model
with $\Omega=0.3$ also fits the rich cluster observational data very well, although
the model with $\Omega=0.5$ clearly does not (even decreasing
$Q_2$ by the maximum allowable 10\% results in only marginal
improvement).  Since we do not observe all possible low-mass groups, the conclusions are
less clear for poorer clusters.  Numerical simulations of these models are needed to
include observational selection effects.  Nevertheless, the fact that both the CHDM and
the $\Lambda$+CDM models give similar results for the GMF indicates that our modelling
of the selection effects is not strongly dependent on the assumed CHDM cosmology.

\section{Conclusions}

We have constructed a sample of optically-selected poor galaxy clusters
and measured their X-ray properties from the ROSAT all-sky survey.  This is the most
extensive sample of poor cluster X-ray properties published to date.  We combined this
sample of poor clusters with a complete subsample of $z\le0.15$ Abell clusters with similar RASS
data to form a list of clusters with a wide range of richnesses, masses, and ICM
properties.  In an effort to understand the observational selection effects, we
used a CHDM numerical simulation to model the cluster Gas Mass Function.
Both the observational results (volume density vs. $z$) and the numerical model indicate
that our sample is approximately volume-limited for groups with more than 4 galaxies
brighter than $M^{\star}$ within a radius of 180 $h^{-1}$ kpc.  The difference between the
Press-Schechter and simulated ``observed'' groups at low $M_{gas}$ is due to the lack
of brighter galaxies in some groups. The observed GMF agrees well with a simple
CHDM, flat Universe model with a constant baryon fraction of 7.5\%, although we do not
claim that it is a unique fit to the data.

From our optically-selected samples, we have produced the first X-ray Luminosity Function
that includes both rich and poor
clusters and, thus, spans a wide range in luminosities.  We find that both the XLF and the
intracluster gas mass function are smoothly continuous curves from poor to rich clusters. 
These functions are broadly consistent with a hierarchical clustering model in which rich
clusters are formed via mergers of poorer clusters, and this process is on-going at the
present epoch.  Renzini et al. (1993) have argued that strong stellar winds early in the life
history of ellipticals may result in the ejection of much of the original gas from galaxy
groups.  However, the continuity of the XLF and GMF over a large range in cluster richness
is not consistent with the expulsion of a significant quantity of gas from poor groups.
 
Our new XLF, in particular, should be useful for studies of the evolution of optically-selected
clusters.  Our rich + poor cluster XLF will provide a low $z$ baseline against which higher $z$ cluster XLFs can
be compared.  There has been tantalizing evidence presented for evolution in the cluster XLF
for some samples of $z>$0.3 clusters (e.g., Henry et al. 1992; Castander et al. 1994).  The abundance of poor clusters relative to rich clusters will provide important constraints on detailed models for
the formation of rich clusters (e.g., Castander et al. 1995).

\acknowledgments

This research was supported by NASA grant NAGW-3152 to JOB \& MLN, NSF grant AST-9317596
to JOB \& CL, and NSF grant ASC-9318185 to MLN \& GLB. JOB would like to thank MPE and its director, Dr. J. Tr\"umper, for access to the RASS database and their hospitality during
the initial stages of this work.  We thank J. Primack and J. Holtzman for valuable
discussions, and H. B\"ohringer and R. Mushotzky for comments on the text.

\clearpage

\figcaption{Overlays of RASS X-ray contours onto optical images from the digital POSS-I.
Contour levels are as follows:
2, 3, 4, 6, 8, 10, 12, 14, 16, 18, 20, 22, 22, 24 $\times 6 \times 10^{-6}$
counts/sec/($15\arcsec \times 15 \arcsec$ pixel) for N34-173; 2, 4, 8, 16, 32 $\times
1.7 \times 10^{-5}$ counts/sec/pixel for N67-335; 2, 4, 6, 8, 10, 12, 14,
16, 18, 20 $\times 6.2 \times 10^{-6}$ counts/sec/pixel for N45-389; 2, 3, 4, 6, 8,
10, 12, 15, 20, 25, 30, 40 $\times 8.5 \times 10^{-6}$ counts/sec/pixel for N56-395.
\label{fig1}}

\figcaption{The differential cluster X-ray Luminosity Function.  The open circles are
from our statistical sample of poor clusters (also includes 8 Abell clusters) with
$0.01 \le z \le 0.03$.  The filled circles are from a complete sample of Abell clusters
with $z \le 0.15$.  The error bars are $1\sigma$.  The line is the best fit
to a power-law.  \label{fig2}}

\figcaption{The integral cluster Gas Mass Function for the observed poor clusters
(open circles), the Abell clusters (filled circles), and the  numerically-``observed''
clusters from the CHDM simulation (squares).  We have assumed $h=0.5$ in order to
plot the observationally-determined masses.  Using the same mass bins each time, 500
realizations of the observed GMF were made with randomly applied errors of up to 50\% on the
masses; however, we found that the Poisson error bars (shown here) dominate.  Press-Schechter predictions
for the GMF are shown for the CHDM model (solid line) and flat, $\Lambda$+CDM universes
with $\Omega$=0.5 (dotted line) and $\Omega$=0.3 (dashed line).  \label{fig3}}

\end{document}